\documentclass[twocolumn,twocolappendix]{aastex631}
\usepackage{amsmath}
\usepackage{amssymb}
\usepackage{booktabs}
\usepackage{multirow}
\usepackage{appendix}
\usepackage{graphicx}
\usepackage{url}
\usepackage{color}
\usepackage[T1]{fontenc}

\begin{document}

\title{A Falsifiable Timing Test for the Double-White-Dwarf Model of Long-Period Transients}

\author[0009-0007-6439-6891]{Yejing Zhan$^{\ast}$}
\affiliation{School of Astronomy and Space Science, Nanjing University, Nanjing 210093, People's Republic of China}

\author[0000-0003-1357-7135]{Di Wang$^{\ast}$}
\affiliation{School of Astronomy and Space Science, Nanjing University, Nanjing 210093, People's Republic of China}

\author[0000-0003-4157-7714]{Fa-Yin Wang}
\affiliation{School of Astronomy and Space Science, Nanjing University, Nanjing 210093, People's Republic of China}
\affiliation{Key Laboratory of Modern Astronomy and Astrophysics (Nanjing University), Ministry of Education, Nanjing 210093, People's Republic of China}

\thanks{These authors contributed equally to this work.}

\correspondingauthor{Fa-Yin Wang}
\email{fayinwang@nju.edu.cn}

\begin{abstract}
Long-period transients (LPTs) are a newly identified class of radio sources with burst recurrence times from minutes to hours, and their diversity suggests multiple physical origins. CHIME/ILT J1634+44, with a short period of 841 s, a long-period modulation of 4206 s, and a significant negative period derivative, strongly suggests a binary origin. For such a short-period source, Roche-lobe constraints strongly favor an ultra-compact companion, motivating a double-white-dwarf (WD--WD) interpretation. 
In this Letter, we show that the WD--WD channel makes a sharp timing prediction: if the burst period is the orbital clock and the long-period modulation is a spin-orbit beat, then the modulation period is not a free timescale. Instead, it must evolve jointly with the orbital clock and the spin clock through gravitational-wave losses, magnetic dissipation, and tidal interaction. For CHIME/ILT J1634+44-like parameters, we find that the beat clock drift $|\dot P_b|\sim 10^{-10} \text{ s s}^{-1}$, implying an observed-minus-calculated drift of tens of seconds in one year. Joint measurements of the burst period, modulation period, and their derivatives provide a minimal and falsifiable timing test of an ultra-compact binary origin.
\end{abstract}

\section{Introduction} \label{sec:intro}

Long-period transients (LPTs) are a novel class of radio sources, characterized by coherent, highly polarized bursts at periods ranging from minutes to hours \citep{hurley-walkerLongperiodRadioTransient2023,hurley-walkerRadioTransientUnusually2022,dongCHIMEFastRadio2025,blootStronglyPolarisedRadio2025,wangDetectionXrayEmission2025,hymanPowerfulBurstingRadio2005,mcsweeneyNewLongperiodRadio2025,anumarlapudiASKAPJ144834685644Newly2025,hurley-walker29HrPeriodic2024,deruiterSporadicRadioPulses2025,leeEmissionInterpulses645hperiod2025,pritchardDiscovery36minuteLongperiod2026}. 
Their phenomenology is diverse, including polarization pattern, burst substructure, and, in some cases, secular timing evolution. The diversity indicates that current LPTs may not originate from a single physical channel. 
\par One major hypothesis is that some LPTs originate from ultra-slow-spin pulsars or magnetars, where the radio bursts are produced by pulsar-like magnetospheric activity \citep{hurley-walkerRadioTransientUnusually2022,konarEnigmaGLEAMXJ162759552350432023,reaLongperiodRadioPulsars2024}. This scenario naturally explains the polarization pattern and substructure of the bursts for some LPTs \citep{hurley-walkerLongperiodRadioTransient2023,calebEmissionstateswitchingRadioTransient2024,cooperRotationalDeathlineRadio2024}. However, when applied across the class, it faces two generic difficulties. First, pulsars and magnetars are usually born with fast spins and therefore require specific evolutionary channels, such as accretion fallback or long-term magnetic braking, to reach the minute-to-hour spin periods \citep{ronchiLongperiodPulsarsPossible2022,xuInteractingActiveLifetime2024,fanEvolutionaryOriginUltralongperiod2024}. Second, most LPTs lie in the pulsar death-valley region, where the pulsar can no longer sustain radio-loud emission \citep{reaLongperiodRadioPulsars2024,reaLongPeriodTransients2026}. These issues suggest that at least some LPTs may require an alternative origin.

\par An important alternative channel is provided by white dwarf (WD) binary systems. In this scenario, the radio emission originates from the magnetic interaction within the binary. Joint radio and optical spectroscopic observations have identified some LPTs as the white dwarf-M dwarf binaries \citep{hurley-walker29HrPeriodic2024,deruiterSporadicRadioPulses2025}. Recent work suggests that CHIME/ILT J1634+44 (hereafter J1634) may also have a binary origin.  This source exhibits a short burst period of 841 s, a long-period modulation of 4206 s, and a significant negative period derivative \citep{blootStronglyPolarisedRadio2025,dongCHIMEFastRadio2025}, making it one of the most constraining LPTs for binary models. A detached binary system is particularly attractive in this channel since coherent radio emission is disfavored in a strongly accreting system, and high-energy counterparts (X-ray, $\gamma$-ray) have not been observed in most LPT sources \citep{papittoSwingsRotationAccretion2013,yangMagneticWhiteDwarf2025}. For such a short-period source, the standard WD--M dwarf scenario is therefore strongly limited by the Roche-lobe constraint, and even an ultracool dwarf companion is still challenged in the compact-companion regime \citep{blootStronglyPolarisedRadio2025,suvorovRevealingNatureUltralong2025}. These constraints naturally motivate an ultra-compact companion, leading to a double-white-dwarf (WD--WD) interpretation.

\par In this Letter, we present a timing diagnostic for the compact binary interpretation, motivated by the J1634. The main purpose of this work is to provide timing diagnostics of the WD--WD scenario under minimal assumptions, rather than to construct a detailed radio-emission model. In this scenario, we point out that once the 841 s burst period is identified as the orbital clock, and the 4206 s long-period modulation arises from spin-orbit beat, the modulation period is no longer an independent timescale. It would drift jointly with the orbital and the spin clocks. Therefore, this drift provides falsifiable timing predictions to diagnose the binary origin of the J1634 and other short-period LPT sources.

\par This paper is organized as follows. Sect. \ref{sec:ui-model} briefly introduces minimal assumptions and energy budget. Sect. \ref{sec:modulation} and Sect. \ref{sec:evolution} demonstrates the interpretation of the modulation and its secular evolution in the WD--WD scenario. Sect. \ref{sec:application} shows the application on J1634-like sources. Finally, Sect. \ref{sec:discussion} summarizes our conclusions.

\section{Minimal assumptions and energy budget}\label{sec:ui-model}
We consider a WD--WD binary system consisting of a magnetized WD with a mass of $M_1$, a radius of $R_1$, a magnetic dipole moment of $\mu=BR_1^3$, together with an unmagnetized WD of mass $M_2$, and radius $R_2$. The binary is assumed to move on a circular orbit with a separation of $a$, a period of $P_0$, and an orbital frequency of $\Omega_0=2\pi/P_0$. The spin frequencies of the magnetized primary and the unmagnetized companion are $\Omega_1$ and $\Omega_2$, respectively. The robustness of the timing prediction under more general magnetic configurations is discussed in Sect.~\ref{sec:discussion}.
\par The motion of the unmagnetized companion in the magnetic dipole field induces an electromotive force (EMF), $\mathcal E\sim 2R_2|\mathbf E|$, where $\mathbf E=\mathbf v_{\text{rel}}/c\times\mathbf B(r)$ is the induced electric field, with local magnetic field at distance $r$, $B(r)=\mu/r^3$. Due to the plausible fast spin of the WD companion, the relative velocity contains two components: the orbital-spin asynchronism between the local field and the companion and the surface motion of the companion itself. Therefore, we write
\begin{equation}
    v_\text{rel}\sim a |\Omega_0-\Omega_1|+\eta R_2|\Omega_0-\Omega_2|
\end{equation}
where $\eta\sim 1$ is a geometric factor that depends on the companion spin orientation. For compactness, we parameterize the total relative motion by
\begin{equation}
    \Delta\Omega_{\rm UI}\equiv \frac{v_{\rm rel}}{a}=\alpha\Omega_0,
\end{equation}
where $\alpha$ is defined as effective asynchronism. The induced electric field is $\mathbf E=\mathbf v_{\rm rel}\times \mathbf B/c$, and the EMF can be estimated as \citep{laiDCCIRCUITPOWERED2012,piroMAGNETICINTERACTIONSCOALESCING2012}
\begin{equation}
    \mathcal E\simeq \frac{2\mu R_2}{ca^2}\Delta \Omega_\text{UI}
\end{equation}

\par The EMF forms a circuit between two WDs, which dissipates the system energy at a rate  
\begin{equation}
    \dot E_\text{UI}=\frac{\mathcal E^2}{\mathcal R_\text{tot}}
\label{eq:Ediss}
\end{equation}
where $\mathcal R_\text{tot}$ is the effective circuit resistance, which is constrained by the azimuthal twist of the magnetic flux tube, $\zeta_\phi=16a\Delta\Omega_\text{UI}/(c^2 \mathcal R_\text{tot})$. When $\zeta_\phi>1$, the strong twist is expected to break the tube. Therefore, the effective resistance is constrained within $16a\Delta\Omega_\text{UI}/c^2<\mathcal R_\text{tot}<4\pi/c$, where the upper limit is set by the vacuum impedance (see Appendix \ref{app:resistance}). 
\par Then, the upper and lower limits of the dissipation energy are bounded by
\begin{equation}
    \begin{aligned}
        \dot E_\text{UI,min}=&\frac{\mu^2R^2_2\alpha^2 \Omega_0^2}{\pi ca^4}\simeq 7.27\times 10^{30}\text{erg s}^{-1} \\&~~\times\alpha^2 \mu^2_{34} R^2_{2,0.02} M_\text{tot,1}^{-4/3}P_{0,14}^{-14/3}\\
        \dot E_\text{UI,max}=&\frac{\mu^2R^2_2\alpha \Omega_0}{4 a^5}\simeq 1.72\times 10^{33}\text{erg s}^{-1} \\&~~\times\alpha \mu^2_{34} R^2_{2,0.02} M_\text{tot,1}^{-5/3}P_{0,14}^{-13/3}
    \end{aligned}
\end{equation}
where $\mu_{34}=\mu/(10^{34} \text{G cm}^3)$, $R_{2,0.02}=R_2/0.02R_\odot$, $M_\text{tot,1}=(M_1+M_2)/1M_\odot$, and $P_{0,14}=P_0/14 \min$. 


\par Detailed radiation microphysics is beyond the scope of this Letter and has been discussed extensively elsewhere \citep{wuElectricallyPoweredBinary2002,willesElectroncyclotronMaserEmission2004,yangMagneticWhiteDwarf2025,quMagneticInteractionsWhite2025,ferrarioElectronCyclotronMaser2026}. Here, we assume that the magnetic interaction can generate narrow-beamed radio emission for simplicity. We further assume that the radio emission is produced in the magnetic flux tube near the companion as indicated by \cite{quMagneticInteractionsWhite2025,yangMagneticWhiteDwarf2025,ferrarioElectronCyclotronMaser2026}. Under such minimal assumptions, we focus on the timing properties of the WD--WD system and how they can diagnose the origin of short-period LPTs.

\section{Beat-modulated visibility} \label{sec:modulation}
Some LPTs exhibit long-period modulation and strong intermittency \citep{hurley-walkerLongperiodRadioTransient2023,deruiterSporadicRadioPulses2025,blootStronglyPolarisedRadio2025}. In the binary scenario, such behavior can arise naturally from coupling between the orbital motion and the stellar spin. Here, we present an illustrative geometry in a WD--WD system to show that the beat clock can produce a long-period modulation. 


\par The purpose of this section is only to establish the existence of such a geometrical mapping. The main timing diagnostic developed in Sect.~\ref{sec:evolution} does not depend on the uniqueness of the beam geometry or on a detailed emission model.

\par We assume that the modulation originates from the beat between the primary spin frequency $\Omega_1$ and the orbital frequency $\Omega_0$. The corresponding beat frequency is 
\begin{equation}
    \Omega_b=|\Omega_1-\Omega_0|
    \label{eq:beat_freq}
\end{equation}

\par We emphasize that $\Omega_b$ has a different physical meaning from the relative angular frequency $\Delta\Omega_\text{UI}$ introduced in Sect.~\ref{sec:ui-model}. The former describes the evolution of the relative position between the field and the companion, which does not relate to $\Omega_2$, and thus controls the burst visibility, whereas the latter characterizes the local relative motion in the unipolar-inductor circuit and mainly determines the induced EMF and dissipation power.

\par For an inclined magnetic dipole with an orientation of $(\theta_1,\phi_1)$, the magnetic field at the location of the companion $(r,\theta_0,\phi_0)$ is

\begin{equation}
  \begin{aligned}
    B_r&=\frac{2\mu}{r^3}\big[\sin\theta_0\sin\theta_1\cos\phi_b+\cos\theta_0\cos \theta_1\big]\\
    B_\theta&=\frac{\mu}{r^3}\big[\sin\theta_0\cos\theta_1-\cos\theta_0 \sin \theta_1\cos\phi_b\big]\\
    B_\phi&=\frac{\mu}{r^3}\sin\theta_1\sin\phi_b
  \end{aligned}
\end{equation}
where $\phi_b=\phi_1-\phi_0=\Omega_bt$ is the beat phase. Although the dipole field generally has a non-zero azimuthal component, $B_\phi$, we approximate the emission direction within the meridional ($r$-$\theta$) plane, since the relative velocity $\mathbf{v}_\text{rel}=v_\text{rel}\hat{\phi}$ induces an electric field, $\mathbf{E}=(-vB_\theta,\,vB_r,\,0)/c$ without azimuthal component. 
Under this approximation, the local emission direction is fully specified by the projected field-line tangent vector $\hat{\boldsymbol{\tau}}(\Delta  \theta_{\rm t})$ in the meridional plane, giving
\begin{equation}
   \tan \Delta \theta_{\rm t}(\phi_b)=\frac{B_\theta}{B_r}.
   \label{eq:Bline}
\end{equation}

\begin{figure*}[htbp]
\centering
\begin{minipage}[c]{0.32\textwidth}
    \centering
    \includegraphics[width=\linewidth]{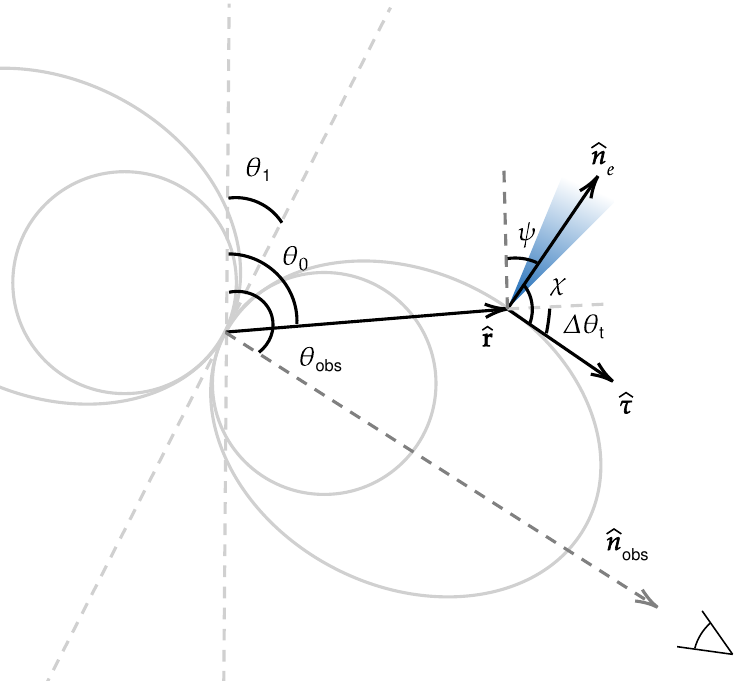}
    {\small (a)}
\end{minipage}
\hfill
\begin{minipage}[c]{0.64\textwidth}
    \centering
    \includegraphics[width=\linewidth]{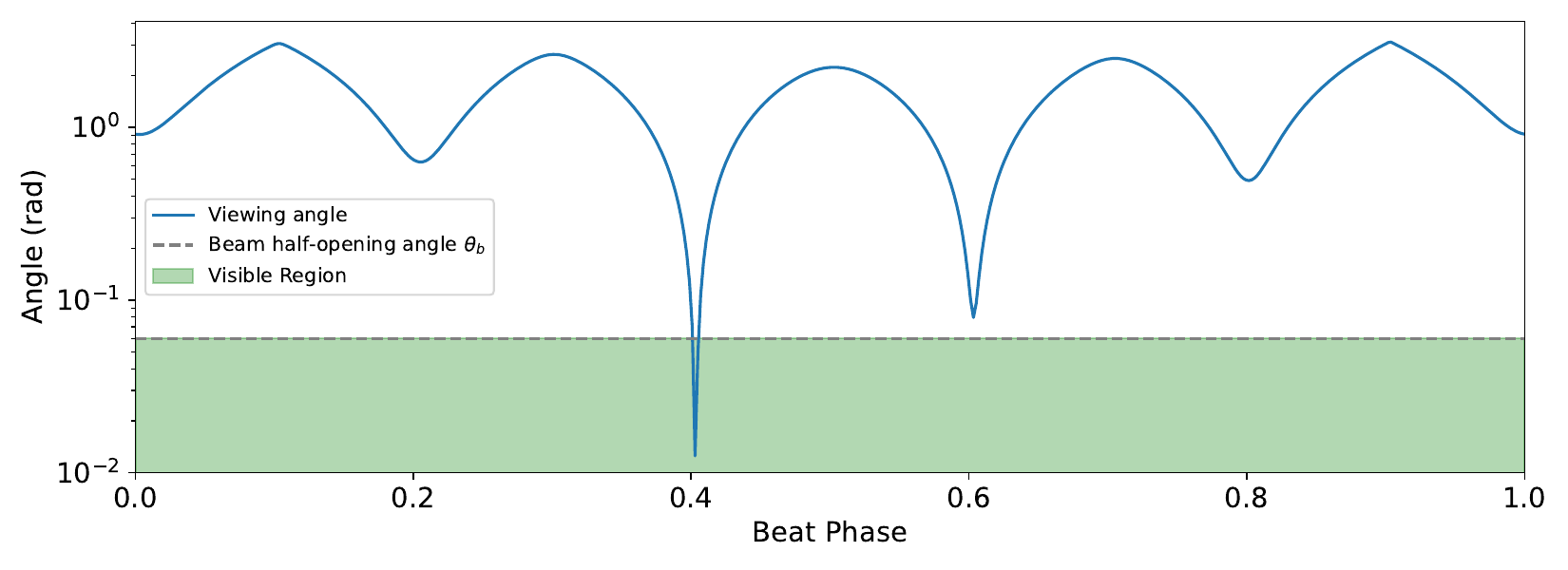}
    {\small (b)}
    
    
\end{minipage}
\caption{(a) Schematic illustration of the emission geometry. The emission beam is centered on $\hat {\boldsymbol n}_e$, which is characterized by the specific angle $\chi$. $\hat{\boldsymbol{n}}_\text{obs}$ denotes the observer’s line of sight.   (b) A fiducial example, motivated by J1634-like timing properties, showing the viewing angle $\mathrm {arccos}(\hat{\boldsymbol{n}}_e\cdot \hat{\boldsymbol{n}}_\text{obs})$ as a function of beat phase. The green shaded region represents the visible condition, i.e., $\mathrm {arccos}(\hat{\boldsymbol{n}}_e\cdot \hat{\boldsymbol{n}}_\text{obs})<\theta_\text{beam}$, where $\theta_\text{beam}$ is the beam half-opening angle. In this case, only one minimum lies within the visibility window during each beat cycle, producing a single-burst pattern. 
}
\label{fig:combo}   
\end{figure*}

As demonstrated in Fig. \ref{fig:combo}(a), we assume that the emission direction $\hat{\mathbf{n}}_\text{e}$ is determined by the local field geometry at the position of the companion. Thus, we define the polar angle of the emission direction to be
\begin{equation}
    \psi=\theta_0+\Delta\theta_t-\chi
\end{equation}
where the term $\chi$ is the angle between the emission direction and the local tangent of the field line, namely $\chi=\text{arccos}(\hat{\boldsymbol{\tau}}\cdot \hat{\mathbf n}_{\rm e})$. 

\par Although the detailed radiation mechanism is uncertain, the following description does not rely on the emission microphysics in detail. We only assume that the local magnetic geometry determines the beam direction through a phenomenological angle $\chi$, whose specific value depends on the radiation model. In this work, we adopt $\chi=\pi/2$ as a fiducial electron-cyclotron-maser-instability-like prescription. 
Different choices of $\chi$ modify the detailed beam geometry but do not alter the existence of beat-modulated visibility.

\par LPTs typically exhibit very small duty cycles, $\Delta t/P\sim10^{-3}$--$10^{-2}$ \citep{mcsweeneyNewLongperiodRadio2025}, implying a narrowly beamed emission pattern. The corresponding beam opening angle can be estimated by the duty cycle,
\begin{equation}
    \theta_\text{beam}=\frac{2\pi \Delta t}{P}\simeq 6\times 10^{-2} \text{ rad} \left(\frac{\Delta t/P}{0.01}\right)
\end{equation}

\par A burst is visible only when the observer’s line of sight lies inside the emission beam. For an observer located at $\hat{\mathbf{n}}_\text{obs}=(\sin\theta_\text{obs}\cos\phi_\text{obs}, \sin\theta_\text{obs}\sin\phi_\text{obs}, \cos\theta_\text{obs})$, the visible condition becomes 
\begin{equation}
    \text{viewing angle}=\text{arccos} [\hat{\mathbf{n}}_e(\phi_b)\cdot \hat{\mathbf{n}}_\text{obs}]<\theta_\text{beam}
\end{equation}
The viewing angle therefore varies over the beat, which indicates that the observer can intercept the emission beam only during restricted beat phase intervals. This naturally produces both the long-period modulation and the strong intermittency observed in partial LPT sources.

\par As an illustrative example, Fig. \ref{fig:combo}(b) shows a fiducial case, motivated by J1634-like timing parameters, with $\Omega_b=1/5\Omega_0$, and we adopt $\theta_0=\pi/2$ and $\theta_1=\pi/6$. In this case, the burst period is highly filtered by the beat and only one burst is visible per beat, resulting in a single-burst pattern. A small drift in the field configuration can change the number of visible minima and can produce a transition between single-burst pattern and double-burst pattern (see Appendix \ref{app:transition}). We present this construction only as an existence proof, while the central timing prediction of this work is the coupled secular evolution of the clocks in the following section.


\section{Beat evolution}\label{sec:evolution}
In a WD--WD binary, both the orbit and the stellar spins evolve under gravitational, magnetic, and tidal torques. Based on the relation, $\Omega_b=|\Omega_0-\Omega_1|$, any evolution on $\Omega_1$ and $\Omega_0$ drives an evolution on the beat. This is a key timing prediction of the WD--WD scenario.    

\par The orbital energy is unavoidably dissipated by the gravitational-wave (GW) emission, which drives a secular orbital decay \citep{petersGravitationalRadiationMotion1964}. The corresponding orbital torque can be written as

\begin{equation}
\begin{aligned}
    N_\text{GW}&=-\frac{32}{5}(2\pi)^{7/3} \frac{G^{7/3}}{c^5}\mathcal M_c^{10/3}P^{-7/3}\\
\end{aligned}
    \label{eq:tor-gw}
\end{equation}
where $\mathcal M_c=(M_1M_2)^{3/5}/(M_1+M_2)^{1/5}$ is the chirp mass.
The unipolar-inductor circuit extracts the orbital energy and exerts a braking torque on the orbit, which can be expressed as
\begin{equation}
    \begin{aligned}
        N_\text{UI}&=-\frac{\dot E_\text{UI}}{\Delta\Omega_\text{UI}}\\
    \end{aligned}
\end{equation}
where $\Delta \Omega_\text{UI}$ is the relative angular frequency defined in Sect. \ref{sec:ui-model}.
The tidal effect provides an additional dissipation of the orbital energy. Following \cite{fullerDynamicalTidesCompact2012}, the tidal torque acting on the $i$th WD can be estimated as
\begin{equation}
    N_{\text{tidal},i}=\frac{3G M_j^2 R_i^5}{2a^6} \frac{k_2}{Q_i} \text{sgn}(\Omega_i-\Omega_0)
\end{equation}
where $i,j\in\{1,2\}$ with $i\neq j$, $k_2\simeq 0.1$ is the Love number of the WD, and $Q_i$ represents the tidal quality factor.


The coupled orbital-spin evolution can therefore be described by 
\begin{equation}
\begin{aligned}
    -\frac{1}{3}\kappa a^2\dot \Omega_0&=N_\text{GW}+N_\text{tidal,1}+N_\text{tidal,2}+N_\text{UI}\\
    I_1\dot \Omega_1&=-N_\text{tidal,1}\\
    I_2\dot \Omega_2&=-N_\text{tidal,2}-N_\text{UI}\\
\end{aligned}
\label{eq:orb-spin_evolution}
\end{equation} 
where $\kappa=M_1M_2/(M_1+M_2)$ is the reduced mass,  and $I_i=\zeta_iM_iR_i^2$ is the rotational inertia of the WD, with the moment of inertia factor $\zeta_i=0.1939(1.44885-M_i/M_\odot)^{0.1917}$ \citep{marshMassTransferDouble2004}. 

 
\par According to the beat relation $\dot \Omega_b=|\dot \Omega_1-\dot\Omega_0|$, the beat evolution is given by 
\begin{equation}
    \dot\Omega_b =
    \begin{cases}
        \dot\Omega_1-\dot\Omega_0, & \Omega_1>\Omega_0,\\
        \dot\Omega_0-\dot\Omega_1, & \Omega_0>\Omega_1.
    \end{cases}
    \label{eq:beat_evo}
\end{equation}

\par Equivalently, given $\dot \Omega=-2\pi/P^2 \dot P$, Eq. \eqref{eq:beat_evo} can be expressed by $\dot P_0, \dot P_1$ and $\dot P_b$,
\begin{equation}
    \dot P_b =
    \begin{cases}
    -\beta^2\dot P_0+(\beta+1)^2\dot P_1, & \Omega_1>\Omega_0,\\
        \beta^2\dot P_0-(\beta-1)^2\dot P_1, & \Omega_0>\Omega_1.
    \end{cases}
    \label{eq:beat_P_evo}
\end{equation}
where $\beta=P_b/P_0$. Eq. \eqref{eq:beat_P_evo} indicates that once both the orbital and beat clocks are identified, the modulation period must evolve jointly with the orbital and spin clocks in a linear relation. 
\par Thus, the beat modulation in a WD--WD system is not only expected, but should also evolve secularly. Alternative isolated-star scenarios may also produce long-term modulation in principle arising from the precession, magnetospheric state changes, or other internal variability. However, unlike the binary cases, such models are hard to provide a stable secular drift on modulation and do not naturally predict a simple algebraic relation between the short period, the modulation period, and their derivatives. In a less compact WD binary, secular period evolution may also exist, but it is barely observable due to a larger binary separation limited by the Roche-lobe. Therefore, a measurement of $P_b$ together with a secular $\dot P_b$ consistent with Eq. \eqref{eq:beat_P_evo} would be difficult to reproduce in these alternative scenarios without additional fine-tuning, constituting especially strong evidence for a compact binary origin. 



\section{Modeling a J1634-like source}\label{sec:application}
Adopting J1634 as a prototypical short-period source, we show that the WD-WD interpretation generically predicts a measurable secular beat drift. Observations show that J1634 has a burst period of 841 s with a long-period modulation of 4206 s and a negative period derivative. Its emission pattern is characterized by the appearance of one or two bursts within each modulation period \citep{blootStronglyPolarisedRadio2025,dongCHIMEFastRadio2025}. Optical observations of J1634 suggest the primary mass $M_1\simeq 0.8M_\odot$ \citep{blootStronglyPolarisedRadio2025}. We assume a fiducial secondary mass of $M_2=0.2 M_\odot$ as a conservative low-mass case, close to the lower-mass boundary of a WD \citep{brownELMSURVEYCOMPLETE2010}. A more massive WD companion is also possible, but the companion mass is constrained by both the observed burst-period derivative and the energy budget. If $M_2$ is too large, the predicted $\dot P_0$ becomes too strong. At the same time, a more massive WD has a smaller radius \citep{zalameaWhiteDwarfsStripped2010}, reducing the unipolar-inductor power because $\dot E_{\rm UI}\propto R_2^2$. These considerations favor a companion mass in the approximate range $M_2 \simeq 0.2$-$0.4 M_\odot$ for J1634-like parameters.

\par We assume that the 841 s burst period is determined by the orbital clock, namely $P_0=2\pi/\Omega_0=841\text{ s}$, and the modulation period arises from the beat clock, $P_b=4206\text{ s}$. According to Eq. \eqref{eq:beat_freq}, the spin period of the primary WD is therefore 
\begin{equation}
    P_1=\frac{1}{P_0^{-1}\pm P_b^{-1}}=\left\{\begin{aligned}
        11.7\text{ min} & \quad\quad \Omega_1>\Omega_0\\
        17.5\text{ min} & \quad\quad \Omega_0>\Omega_1
    \end{aligned}\right.
\end{equation}

\par Both solutions are plausible for a magnetic WD. Moreover, the beat relation fixes the primary spin-orbit offset to be $|\Omega_1-\Omega_0|=\Omega_b\sim 0.2\Omega_0$, which also contributes to the effective asynchronism relevant for the unipolar-inductor circuit.
\par The observed radio luminosity of each burst is $2.3-52\times 10^{29}\text{ erg s}^{-1}$. As we showed in Sect. \ref{sec:ui-model}, such a luminosity can be accommodated within the energy budget of the unipolar-inductor circuit of $0.3-367\times 10^{30}\text{ erg s}^{-1}$, provided that the magnetic dipole moment $\mu=10^{34}\mathrm G~\mathrm{cm}^{3}$ and the effective asynchronous with $\alpha \sim 0.2$. The effective asynchronism is contributed by the spin-orbit offset $|\Omega_0-\Omega_1|\simeq 0.2 \Omega_0$, while the companion remains approximately synchronized $|\Omega_2|\sim \Omega_0$. Therefore, the basic energetics of J1634 can be naturally reproduced in the WD--WD binary framework.

\par The observed transition between single-burst and double-burst patterns can also be accommodated naturally within the beat-modulated visibility framework. In our fiducial geometry, a small drift in the effective local field orientation is sufficient to change the number of visible minima within one beat cycle, as shown in Appendix \ref{app:transition}. We stress that the robust prediction of secular evolution of the beat clock does not rely on the detailed burst morphology. 

\par Using Eq. \eqref{eq:orb-spin_evolution}, we estimate the coupled orbital-spin evolution for J1634-like parameters to be 
\begin{equation}
    \begin{aligned}
        \dot \Omega_0&\sim 7.0\times 10^{-17} \text{ rad s}^{-2}\\
        \dot \Omega_1&\sim  \pm 2.4\times 10^{-17} \text{ rad s}^{-2}
    \end{aligned}
\end{equation}
where the sign of $\dot \Omega_1$ is determined by $\mathrm{sgn}(\Omega_0-\Omega_1)$. In particular, the $\dot \Omega_0$ is dominated by the GW torque $N_\text{GW}$, and the $\dot\Omega_1$ is determined by the primary tidal torque $N_\text{tidal,1}$. The companion spin evolution is unimportant for the secular timing behavior that we discussed, and it is negligible over secular timescales. 

\par Using the relation $\dot \Omega=-2\pi/P^2\dot P$ and Eq. \eqref{eq:beat_evo}, the evolution of the clocks are
\begin{equation}
    \begin{aligned}
        \dot P_0&\sim -7.7\times 10^{-12} \text{ s s}^{-1}\\
        \dot P_b&\sim \begin{cases}
            2.6\times 10^{-10}  \text{ s s}^{-1}& \Omega_1>\Omega_0\\
            -1.3\times 10^{-10}  \text{ s s}^{-1}& \Omega_0>\Omega_1\\
        \end{cases}    
        \\
    \end{aligned}
\end{equation}

\par The predicted orbital decay is consistent with the measured burst-period derivative of J1634, $(-9\pm 3)\times 10^{-12} \text{s s}^{-1}$. More importantly, the model predicts that the beat clock evolves at a faster rate, with $|\dot P_b|$ exceeding $|\dot P_0|$ by about an order of magnitude.

\begin{figure}[h]
\centering
\begin{minipage}[c]{1\linewidth}
    \centering
    \includegraphics[width=\linewidth]{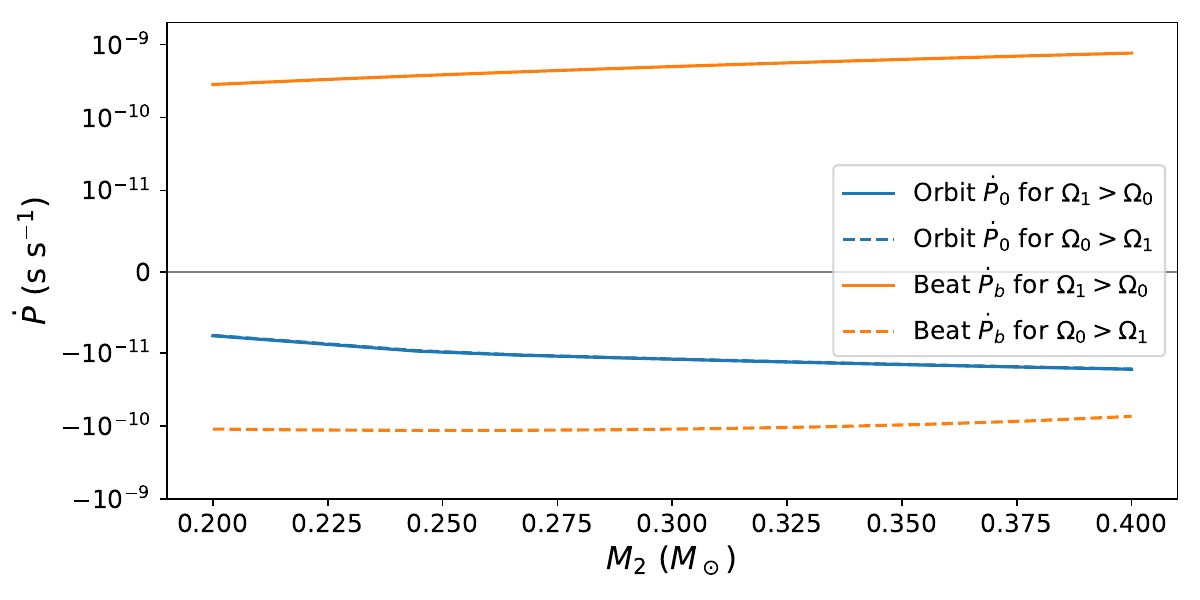}
    {\small (a)}
\end{minipage}

\begin{minipage}[c]{1\linewidth}
    \centering
    \includegraphics[width=\linewidth]{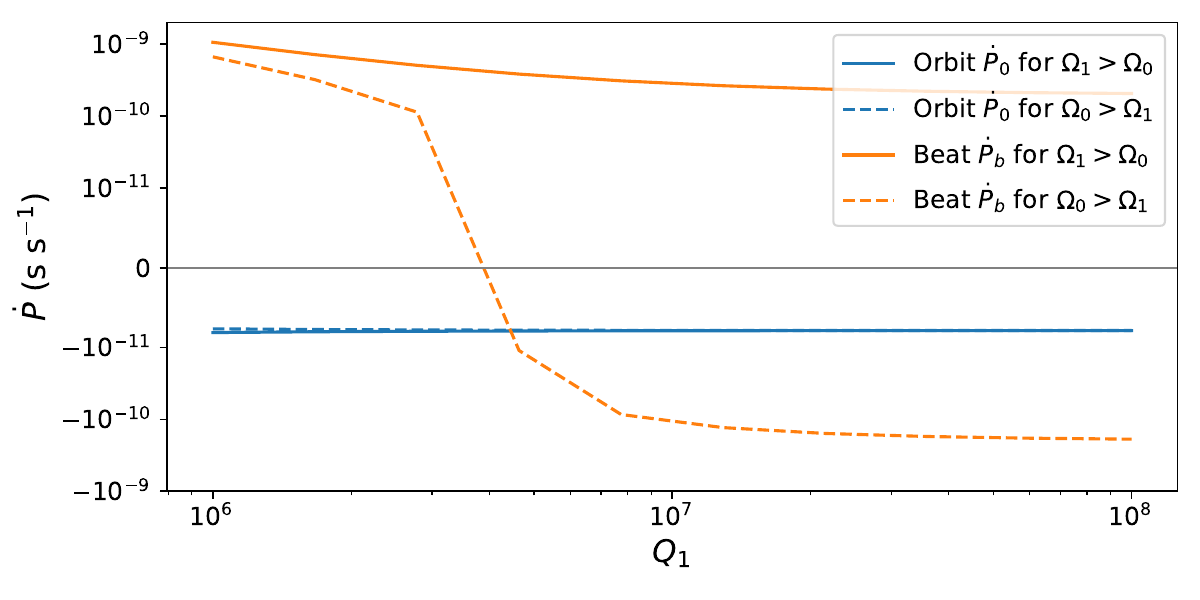}
    {\small (b)}
\end{minipage}
\caption{The dependence of the period evolution on the secondary mass $M_2$ and the primary tidal factor $Q_1$.}
\label{fig:scan}   
\end{figure}

\par To test the robustness of our results, we scan the secondary mass $M_2$ and the primary tidal quality factor $Q_1$, whose fiducial values are set to $M_2=0.2M_\odot$ and $Q_1=10^7$ in the previous text. We focus on $Q_1$ rather than the secondary asynchronism $\alpha$ or the secondary tidal quality factor $Q_2$, since the latter parameters have a weaker influence on the orbital and beat evolution and do not qualitatively alter the results, according to Eqs. \eqref{eq:orb-spin_evolution}.

\par As shown in Fig.~\ref{fig:scan}(a), the orbital decay becomes slightly stronger with increasing $M_2$, reflecting the enhanced gravitational-wave losses in a more massive binary. However, over the range $M_2=0.2$--$0.4\,M_\odot$, $\dot P_0$ remains at the level of $\sim10^{-11}\,\mathrm{s\,s^{-1}}$. Over the same range, the beat-period derivative also remains robust, with $|\dot P_b| \sim 10^{-10}$--$10^{-9}\,\mathrm{s\,s^{-1}}$. Figure~\ref{fig:scan}(b) shows that $\dot P_0$ is nearly independent of $Q_1$, again because the orbital evolution is dominated by gravitational-wave losses. In contrast, $\dot P_b$ depends on $Q_1$ through the evolution of the primary spin $\Omega_1$. Across the plausible range $Q_1=10^6$--$10^8$, $|\dot P_b|$ remains of order $\sim10^{-10}$--$10^{-9}\,\mathrm{s\,s^{-1}}$, except in a narrow interval around $Q_1\sim4\times10^6$, where it approaches zero. Thus, the selection of the fiducial $Q_1$, $Q_2$, and $M_2$ introduces only a modest bias in secular beat evolution and orbital evolution.

\par Observationally, the detectability of a period derivative is controlled by the corresponding observed-minus-calculated (O--C) drift,
 \begin{equation}
    \Delta t_\text{O--C}=\frac{1}{2} \frac{|\dot P|}{P}T^2 
\end{equation}
where $P$ is the period and $T$ is the observation baseline (see Appendix \ref{app:O--C}). For the two fiducial branches above, the predicted $|\dot P_b|$ implies a timing O--C drift of $\sim 10-30\text{ s}$ after 1 year and $\sim 100-300\text{ s}$ after 3 years. 
Thus, near-term monitoring of J1634 may be sufficient to test for beat-period evolution and thereby constrain the binary interpretation.

\par The clock evolution provides a distinctive prediction for the WD--WD interpretation. In particular, in the J1634 case, the source should show not only a stable negative $\dot P_0$, but also a measurable secular drift on the modulation period $\dot P_b$. If future monitoring confirms both secular evolution consistent with Eq. \eqref{eq:beat_P_evo}, it would constitute strong evidence in favor of the WD-WD binary scenario.





\section{Summary and discussion}\label{sec:discussion}

In this Letter, we present a timing diagnostic for the WD--WD binary under minimal assumptions and apply this framework to interpret J1634-like short-period LPTs. This framework provides a sharp timing prediction: if the 841 s burst period is the orbital clock and the 4206 s long-period modulation is a spin-orbit beat, then the modulation period is not a free timescale. It must undergo coupled secular evolution together with the orbit and the primary spin. This turns the timing behavior into a quantitatively testable probe of the binary nature and compactness of the system.  

\par Using J1634 as an illustrative example, we find $\dot P_0\sim -10^{-11}\,\mathrm{s\,s^{-1}}$ and $\dot P_b\sim 10^{-10}\,\mathrm{s\,s^{-1}}$. Across representative companion masses and primary tidal quality factors, this coupled evolution remains robust except for a narrow window of $Q_1$, resulting in $\dot P_b$ cancellation. Observationally, the predicted O--C drift of $\dot P_b$ reaches tens of seconds within about a year and hundreds of seconds over a few years, making the test feasible on near-term monitoring baselines. Alternative scenarios, such as isolated-star scenarios and less-compact binary scenarios, are hard to provide a measurable, coupled, and stable drift. Thus, joint measurements of the burst period, modulation period, and their secular drifts can therefore distinguish the compact-binary channel from alternative scenarios.

\par In addition, the WD--WD interpretation is expected to provide a more stable radio-emission environment than a WD--M-dwarf system on long timescales. In a WD--WD binary, the local magnetic field configuration is expected to remain stable due to the clean environment and the absence of strong stellar activity, although it would be modulated by the orbital motion and the spin–orbit beat on a short timescale. In contrast, an M dwarf exhibits strong stellar activity and a complex magnetosphere \citep{yangFlaringActivityDwarfs2017,kochukhovMagneticFieldsDwarfs2020}, which can perturb the local field structure and plasma density, introducing stronger variability and a less stable maser environment.

\par We also note that the timing prediction is developed under the assumption of a dipole magnetic field and an unmagnetized companion. These assumptions are not essential to the timing prediction. A non-dipolar field can change the local beam direction and the visibility window, but, higher-order multipole components decay faster with distance than the dipole component. At the orbital separation, the local magnetic field can therefore be expected to be dipole-dominated, and a non-dipolar field should not substantially change the main timing prediction. If the two WDs carry comparable magnetic moments, direct magnetic reconnection may dominate over the unipolar-inductor mechanism. Such a double-magnetic case can change the dissipation power and beaming geometry \citep{mostElectromagneticPrecursorsGravitationalwave2020,mahlmannElectrodynamicsDissipationBinary2025}, and may introduce additional modulations. However, the central timing prediction still follows from identifying the burst period with the orbital clock and the long modulation period with a beat clock. The additional modulation would then act as a perturbation to the basic timing relation. A full magnetospheric treatment of this case is beyond the scope of this Letter.


\par A particularly important prediction of the WD--WD scenario is persistent mHz gravitational-wave emission. Detection of a GW counterpart from the same sky position, with $f_\text{GW}=2/P$ and a chirp mass or strain amplitude consistent with a WD--WD system, would provide strong evidence for a compact binary origin of the source. For fiducial parameters, the GW counterpart is expected to be detectable with a high signal-to-noise ratio by a space-based detector (see Appendix~\ref{app:gw} for detailed calculations), providing an independent probe of the spin--orbit coupling and tidal effects in the compact binary \citep{fiaccoUncoveringStealthBias2024,ebadiLISADoubleWhite2025}. In this sense, short-period LPTs may serve as an electromagnetic gateway to the ultra-compact WD binary population and as promising multimessenger targets for future space-based GW observatories.

\section*{Acknowledgements}
We thank the anonymous referee for the helpful comments. This work was supported by the National Natural Science
Foundation of China (grant Nos. 12494575 and 12273009).

\appendix
\section{Effective resistance}\label{app:resistance}
In the unipolar-inductor picture, the circuit dissipation is controlled by the effective resistance, $\mathcal R_\text{tot}$. Generally, the resistances from the magnetosphere and the binary stars contribute to the effective resistance, which is denoted by $\mathcal R_\text{space}, \mathcal R_1$, and $\mathcal R_2$ respectively. In a schematic decomposition, one may write
\begin{equation}
    \mathcal R_\text{tot}\sim \mathcal R_1+\mathcal R_2+2\mathcal R_\text{space}
\end{equation}

Since the fractional effective area of the magnetic poles on the surface of the primary WD is significantly small, which indicates $\mathcal{R}_1\gg \mathcal{R}_2$ \citep{wuElectricallyPoweredBinary2002}. The resistance of the primary WD is given by \citep{wuElectricallyPoweredBinary2002,piroMAGNETICINTERACTIONSCOALESCING2012},
\begin{equation}
    \mathcal R_1\simeq \frac{1}{2\sigma_1}  \left(\frac{H}{\Delta d}\right) \left(\frac{a}{R_1}\right)^{3/2} \frac{\mathcal J(e)}{R_2}
\end{equation}
where $\mathcal J\sim 1$ is a geometric factor, and $H/\Delta d\sim 1$ is the ratio of the atmospheric depth factor and the thickness of the dissipative surface layers of the WD. The $\sigma_1\sim 10^{14} \text{ s}^{-1}$ is the conductivity of the surface of the primary. 

\par The resistance of the magnetosphere can be commonly represented in units of the vacuum impedance 
\begin{equation}
    \mathcal R_\text{space}=\frac{4\pi}{c}\eta_s
\end{equation}
where $\eta_s$ is a correction factor, representing the resistance reduction of the effective impedance by the plasma injection. In general, one expects $\eta_s\lesssim 1$.

\par Because the detailed current closure and plasma loading are uncertain, we do not attempt a precise decomposition of $\mathcal R_{\rm tot}$ in the main text. Instead, we use two robust bounds. The lower bound follows from the requirement that the azimuthal twist of the flux tube, $\zeta_\phi=16a\Delta\Omega_{\rm UI}/(c^2\mathcal R_{\rm tot})$. When $\zeta_\phi>1$, the strong twist would break the magnetic flux tube. Thus, the constraint implies
\begin{equation}
    \mathcal R_{\rm tot}\gtrsim \frac{16a\Delta\Omega_{\rm UI}}{c^2}.
\end{equation}
For the upper bound, we adopt the vacuum-impedance scale as a conservative benchmark,
\begin{equation}
    \mathcal R_{\rm tot}\lesssim \frac{4\pi}{c}.
\end{equation}
Therefore, the effective circuit resistance is bracketed by
\begin{equation}
    \frac{16a\Delta\Omega_{\rm UI}}{c^2}\lesssim \mathcal R_{\rm tot}\lesssim \frac{4\pi}{c}.
\end{equation}
These bounds are the only ingredients required for the energetics estimates in the main text.

\section{Transition of the burst pattern}\label{app:transition}

\begin{figure}[h]
\centering
\begin{minipage}[c]{1\linewidth}
    \centering
    \includegraphics[width=\linewidth]{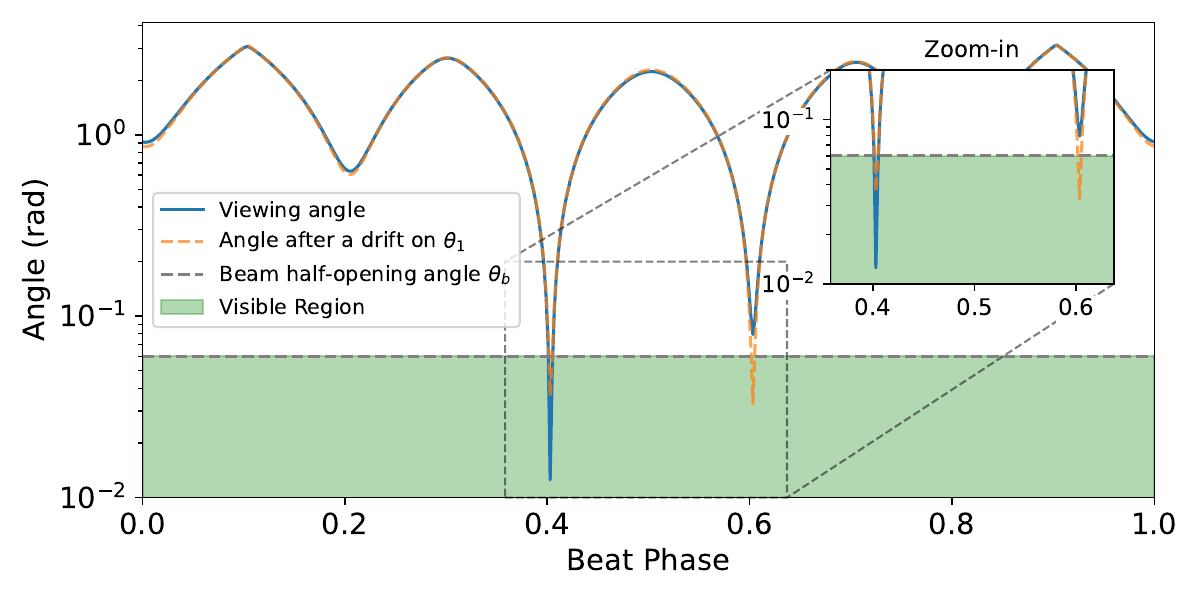}
    {\small (a)}
\end{minipage}

\begin{minipage}[c]{1\linewidth}
    \centering
    \includegraphics[width=\linewidth]{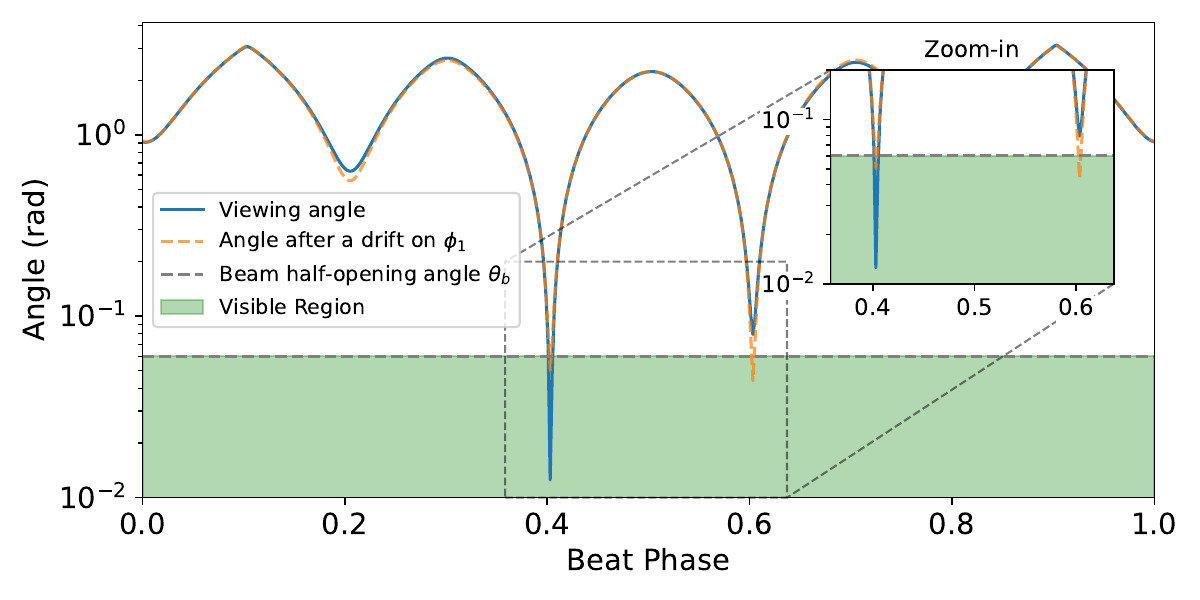}
    {\small (b)}
\end{minipage}
\caption{(a) Same as Fig. \ref{fig:combo}(b), but with an illustrative few-degree-order drift on the magnetic dipole inclination $\theta_1$, marked in orange dashed curve. The inset shows a zoom-in view, highlighting a specific window marked in a dashed box. (b) Same as panel (a), but with an illustrative few-degree-order drift on the magnetic dipole initial azimuth $\phi_1$. Both drifts allow two minima to satisfy the viewing condition, leading to a transition from a single-burst to a double-burst pattern.
}
\label{fig:transition}   
\end{figure}

\par Due to the sensitive dependence of the viewing angle on the local magnetic geometry, even a small drift in the field configuration can change the burst pattern. Rather than specifying the physical origin of the drift, we show that a drift of order a few degrees in the effective local field orientation, as an illustrative example, is sufficient in our fiducial geometry to change the number of visible minima per beat cycle. As demonstrated in Fig. \ref{fig:transition}, the drifts on the inclination and the azimuth of the magnetic dipole allow two minima to satisfy the viewing condition, leading to a transition from a single-burst to a double-
burst pattern.

\section{O--C drift analysis}\label{app:O--C}
In standard timing analysis, the burst phase can be expanded as \citep{hobbsTempo2NewPulsartiming2006,luoPINTModernSoftware2021} 
\begin{equation}
    \Phi(T)=\Phi_0+\nu T+\frac{1}{2}\dot \nu T^2 +\mathcal O(T^3)
    \label{eq:phase-taylor}
\end{equation}
where $\nu=1/P$ is the burst frequency with the burst period $P$, and $\dot\nu=-1/P^2 \dot P$ is the burst frequency derivative. If the source is fitted with a constant period, the quadratic term would remain, leaving a phase residual,
\begin{equation}
    \Delta \Phi(T)\simeq\frac{1}{2} \dot\nu T^2
\end{equation}

The corresponding observed-minus-calculated (O–C) time residual is therefore
\begin{equation}
    \Delta t_\text{O--C}=|P\Delta\Phi(t)|\simeq \frac{1}{2} \frac{|\dot P|}{P}T^2 
\end{equation}
where $T$ is the observation baseline. 



\section{gravitational wave detection}\label{app:gw}
Ultra-compact WD--WD binary is a main target of the space-based GW detectors, like LISA, Taiji, and Tianqin \citep{amaro-seoaneLaserInterferometerSpace2017,huTaijiProgramSpace2017,luoTianQinSpaceborneGravitational2016}. In the quadruple approximation, the strain amplitude for the GW emission of the WD--WD binary is \citep{robsonConstructionUseLISA2019}
\begin{equation}
    h_0(f)=\frac{4(G\mathcal M_c)^{5/3}(\pi f)^{2/3}}{c^4 D}
\end{equation}
where $D$ is the distance of the source, and $f=f_\text{GW}=2f_\text{orb}$ is the frequency of the GW. 
\par We assume that the orbital frequency evolves negligibly over the observation time $T_\text{obs}$, so that the source can be treated as quasi-monochromatic. Thus, the signal-to-noise ratio of the GW signal is 
\begin{equation}
\begin{aligned}
    \rho\simeq& \sqrt{\frac{N_\text{cyc}h_0^2}{fS_n(f)}}\\
    \simeq &98.1 ~\mathcal M_{c,0.3}^{5/3}P^{-2/3}_{14}D_5^{-1}T_\text{obs,4}^{1/2}S_\text{n,-40}^{-1/2}
\end{aligned}
\end{equation}
where $S_n(f)$ is the one-sided noise spectrum of the detector, and $N_\text{cyc}=fT_\text{obs}$ is the number of GW cycles during the $T_\text{obs}$. Moreover, $D_5=D/(5 \text{ kpc})$, $T_\text{obs,4}=T_\text{obs}/(4\text{ yr})$, and $S_{n,-40}=S_n/10^{-40}\text{Hz}^{-1}$. 

\par Therefore, for a J1634-like ultra-compact WD--WD binary at a distance of a few kpc, the mHz GW counterpart is expected to be detectable with a high signal-to-noise ratio by future space-based GW observatories. Combined with the electromagnetic timing observables, such a detection would offer a powerful multimessenger probe of the binary parameters, spin--orbit coupling, and tidal effects.

\bibliography{Cosmology}{}

\bibliographystyle{aasjournal}

\end{document}